\documentclass[aps,showpacs,twocolumn]{revtex4}
\usepackage{mathrsfs}
\usepackage{amssymb}
\usepackage{amsmath}
\usepackage{bm}
\usepackage{graphics}
\usepackage{natbib}
\newcommand{\llangle}{\langle\!\langle}
\newcommand{\rrangle}{\rangle\!\rangle}

\begin{document}
\title{Transport properties in a Non-Hermitian triple-quantum-dot structure}
\author{Lian-Lian Zhang}
\author{Wei-Jiang Gong}\email{gwj@mail.neu.edu.cn}

\affiliation{College of Sciences, Northeastern University, Shenyang
110819, China}
\date{\today}

\begin{abstract}
In this work, we study the effect of $\mathcal{PT}$-symmetric complex potentials on the transport properties of one non-Hermitian system, which is formed by the coupling between a triple-quantum-dot molecule and two semi-infinite leads. As a result, it is found that the $\mathcal{PT}$-symmetric imaginary potentials take pronounced effects on transport properties of such a system, including changes from antiresonance to resonance, shift of antiresonance, and occurrence of new antiresonance, which are determined by the interdot and dot-lead coupling manners. This study can be helpful in understanding the quantum transport behaviors modified by the $\cal PT$ symmetry in non-Hermitian discrete systems.
\end{abstract}
% \keywords{}
\pacs{73.23.Hk, 73.50.Lw, 85.80.Fi} \maketitle

\bigskip

\section{Introduction}
The systems of non-Hermitian Hamiltonians have opportunities to exhibit entirely real spectra if they possess parity-time ($\mathcal{PT}$) symmetry\cite{Bender1}. This exactly means the fundamental physics intension and potential application of such kinds of systems. Therefore, researchers explored numerous $\mathcal{PT}$-symmetric
systems from various aspects in the past decades, including the complex
extension of quantum mechanics\cite{Bender3,Mosta}, the quantum
field theories and mathematical physics\cite{Bender4}, open quantum
systems\cite{Rotter}, the Anderson models for disorder systems\cite{Gold,Hei,Moli},
the optical systems with complex refractive indices\cite{Klaiman,Xu,Kottos,Makris,Luo}, and the topological insulators\cite{Hu,Zhu}. Besides, the non-Hermitian lattice models with $\mathcal{PT}$ symmetry have attracted much attention, following the experimental achievement in optical waveguides\cite{Guo,Kip}, optical lattices\cite{Miri}, and in a pair of coupled
LRC circuits\cite{Li}. These works make the warming up of the researches about the non-Hermitian Hamiltonians with $\mathcal{PT}$ symmetry. Recently, it has been reported that such systems can be realized in the Gegenbauer-polynomial quantum chain\cite{Zo}, one-dimensional $\mathcal{PT}$-symmetric chain with disorder\cite{Bendix}, the chain model with two conjugated imaginary potentials at two end
sites\cite{Jin}, the tight-binding model with position-dependent
hopping amplitude\cite{Jogle}, and the time-periodic $\mathcal{PT}$-symmetric
lattice model\cite{Valle}.
\par
Accompanied by the exploration and fabrication of the $\cal PT$-symmetric systems of non-Hermitian Hamiltonians, the physics properties of them have also become one important concern in the field of quantum physics. On the one hand, researchers have began to focus on their $\cal PT$ phase diagrams as well as the signatures of $\cal PT$-symmetry breaking. On the other hand, some works paid attention to the quantum transport properties in these systems, and various interesting results have been observed\cite{PRA0}. It has been shown that for a Fano-Anderson system, the $\mathcal{PT}$-symmetric imaginary potentials induce some pronounced effects on Fano interference, including changes from the perfect reflection to perfect transmission, and rich behaviors for the absence or existence of the perfect reflection at one and two resonant frequencies\cite{PRA1}. In a non-Hermitian Aharonov-Bohm ring system with a quantum dot(QD) embedded in each of its two arms, it has been observed that with appropriate parameters, the asymmetric Fano profile will show up in the conductance spectrum just by non-Hermitian quantity in this system\cite{Lvrong}. Thus one can ascertain that in the $\cal PT$-symmetric systems of non-Hermitian Hamiltonians, the $\mathcal{PT}$-symmetric imaginary potentials play nontrivial roles in modulating the quantum interference that governs the quantum transport process. Surely, for further understanding the role of $\mathcal{PT}$-symmetric imaginary potentials in modifying the transport properties in low-dimensional systems, some other typical geometries should be investigated.
\par
In the present work, we would like to study the effect of $\mathcal{PT}$-symmetric complex potentials on the transport properties of non-Hermitian systems, which is formed by the coupling between a triple-QD molecule and two semi-infinite leads. By analytically solving the scattering process, we find that the $\mathcal{PT}$-symmetric imaginary potentials can induce pronounced effects on transport properties of our systems, including changes from antiresonance to resonance, shift of antiresonance, and occurrence of new antiresonance, which are related to the interdot and QD-lead coupling manners. This study can assist to understand the quantum transport behaviors modified by the $\cal PT$ symmetry in non-Hermitian discrete systems.
\begin{figure}
\centering \scalebox{0.40}{\includegraphics{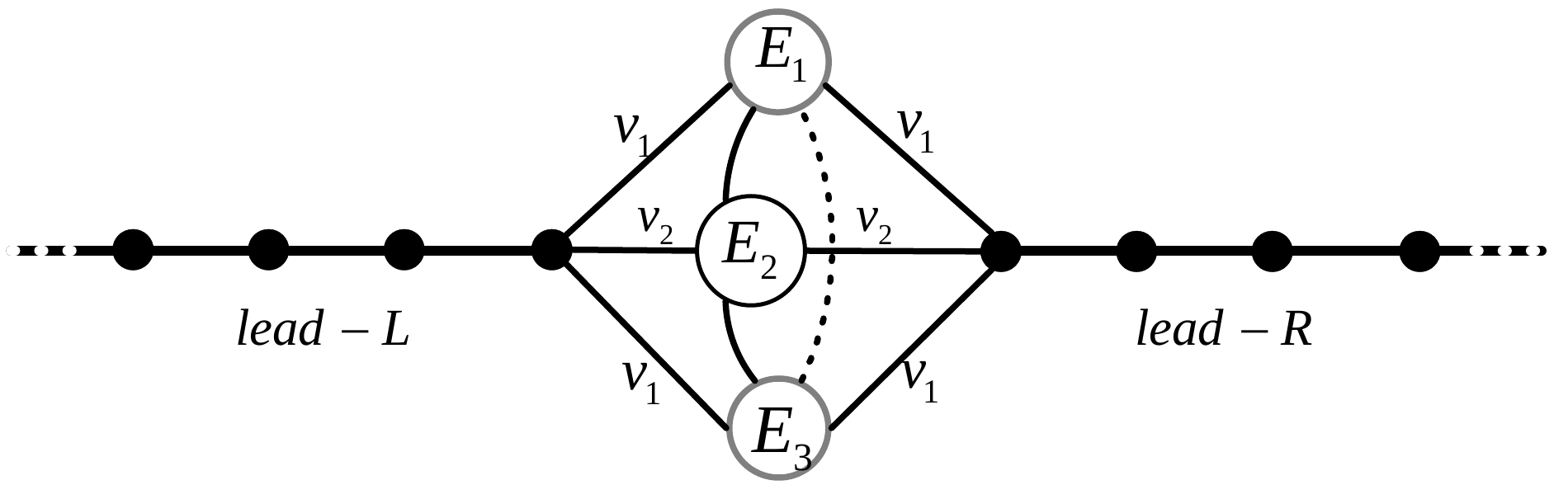}} \caption{
Schematic of one non-Hermitian triple-QD circuit, in which each QD couples to two leads simultaneously. The terminal QDs are influenced by $\cal PT$-symmetric complex on-site chemical potentials. \label{structure}}
\end{figure}
\section{Theoretical model}
The structure that we consider is shown in Fig.\ref{structure}, in which each QD of a triple-QD molecule couples to two metallic leads, respectively. We add imaginary potentials to the two terminal QDs (QD-1 and QD-3) to represent the physical gain or loss during the interacting processes between the environment and it. The Hamiltonian of this system reads
\begin{eqnarray}
H=\sum_\alpha H_\alpha +H_c+\sum_\alpha H_{\alpha T},
\end{eqnarray}
with its each part given by
\begin{eqnarray}
H_\alpha&=&\sum_{j=1}^{\infty}t_{0}c^{\dag}_{\alpha j}c_{\alpha,j+1}+h.c. \notag\\
H_c&=&\sum_{l=1}^3 E_ld_{l}^{\dag}d_{l}+\sum_{l'=1}^{2}t_{l'}d_{l'+1}^{\dag}d_{l'}+t_3d_{1}^{\dag}d_{3}+h.c.,\notag\\
H_{\alpha T}&=&\sum_lv_{\alpha l}c_{\alpha1}^{\dag}d_{l}+h.c..\label{Hamilton}
\end{eqnarray}
$d^\dag_l$ ($d_l$) is the creation (annihilation)
operator for QD-$l$ with energy level $E_l$. When $E_l$ are real, the Hamiltonian
is Hermitian, whereas if one of them is complex,
its Hamiltonian becomes non-Hermitian. $c^\dag_{\alpha j}$
($c_{\alpha j}$) is to create (annihilate) a fermion at site-$j$ of lead-$\alpha$ with $t_0$
being the hopping amplitude between the nearest sites. $v_{\alpha l}$ is the tunneling amplitude between QD-$l$
and lead-$\alpha$. It should be noted that in discrete systems, $\cal P$ and $\cal T$ are defined as the space reflection (parity) operator and the time-reversal operator. A Hamiltonian is said to be $\cal PT$ symmetric if it obeys the commutation relation $[{\cal PT}, H]=0$. With respect to our considered geometry, the effect of the $\cal P$ operator is to let ${\cal P}d_{N+1-l}{\cal P}=d_l$ with the linear chain as the mirror axis, and the effect of
the $\cal T$ operator is ${\cal T} i{\cal T}=-i$. Thus, it is not difficult to find that the
Hamiltonian is invariant under the combined
operation ${\cal PT}$, under the condition of $t_1=t_2$, $v_{\alpha 1}=v_{\alpha'N}$, and $E_l=E^*_{N+1-l}$.
\par
The study about the quantum transport through this
structure depends on the calculation of transmission function in this system. According to the previous works, various methods can be employed to calculate the transmission function. In this work, we would like to choose the nonequilibrium Green function technique to perform the calculation about it. Therefore, the transmission function can directly be expressed as\cite{GF1,GF2}
\begin{equation}
T(\omega)=\mathrm{Tr}[\Gamma^L G^a(\omega)\Gamma^R
G^r(\omega)].\label{conductance}
\end{equation}
$\Gamma^{\alpha}=i(\Sigma_{\alpha}-\Sigma^\dag_{\alpha})$ denotes
coupling between lead-$\alpha$ and the device region. $\Sigma_{\alpha}$, defined as $\Sigma_{jl,\alpha}=v_{\alpha j}v^*_{\alpha l}g_\alpha$, is the selfenergy caused by the coupling between the
quantum chain and lead-$\alpha$. $g_\alpha$ is the Green function of the end site of the semi-infinite lead. Due to the uniform intersite coupling in lead-$\alpha$, the analytical form of $g_\alpha$ can be written out. For the same lead-$L$ and lead-$R$, we can obtain the result that $g_\alpha=g_0={\omega\over 2t_0^2}-i\rho_0$ with $\rho_0={\sqrt{4t_0^2-\omega^2}\over 2t_0^2}$\cite{Self}. Additionally, in Eq.(\ref{conductance}) the
retarded and advanced Green functions in Fourier space are involved.
They are defined as follows:
$G_{jl}^r(t)=-i\theta(t)\langle\{d_{j}(t),d_{l}^\dag\}\rangle$
and
$G_{jl}^a(t)=i\theta(-t)\langle\{d_{j}(t),d_{l}^\dag\}\rangle$,
where $\theta(x)$ is the step function. The Fourier transforms of
the Green functions can be performed via
$G_{jl}^{r(a)}(\omega)=\int^{\infty}_{-\infty}
G_{jl}^{r(a)}(t)e^{i\omega t}dt$. They can be solved by means of the equation of motion method. For
convenience we employ an alternative notation $\llangle
A|B\rrangle^x$ with $x=r,a$ to denote the Green functions
in Fourier space, e.g., $G^r_{jl}(\omega)$ is identical to
$\llangle d_{j}|d^\dag_{l}\rrangle^r$. In general, the
Green functions obey the following equations of motion:
\begin{eqnarray}
(\omega\pm i0^{+})\llangle
A|B\rrangle^{r(a)}=\langle\{A,B\}\rangle+\llangle
[A,H]|B\rrangle^{r(a)}.\label{motion}
\end{eqnarray}
Starting from Eq.(\ref{motion}), we can derive the equation
of motion of the retarded Green function $\llangle
d_{j}|d^\dag_{l}\rrangle^r$ between two arbitrary QDs. And then, the matrix form of the retarded
Green function in this system can be obtained, i.e.,
\begin{small}
\begin{eqnarray}
&&[G^r]^{-1}=\notag\\
&&\left[\begin{array}{ccc} z-E_1-\Sigma_{11} &-t_1^*-\Sigma_{12}&-t_3-\Sigma_{13}\\
-t_1-\Sigma_{21} &z-E_2-\Sigma_{22}& -t_2^*-\Sigma_{23}\\
-t^*_3-\Sigma_{31}&-t_{2}-\Sigma_{32}& z-E_3-\Sigma_{33}\\
\end{array}\right]\ \label{green}
\end{eqnarray}
\end{small}
with $\Sigma=\Sigma_L+\Sigma_R$.
\par

\begin{figure}[htb]
\centering \scalebox{0.38}{\includegraphics{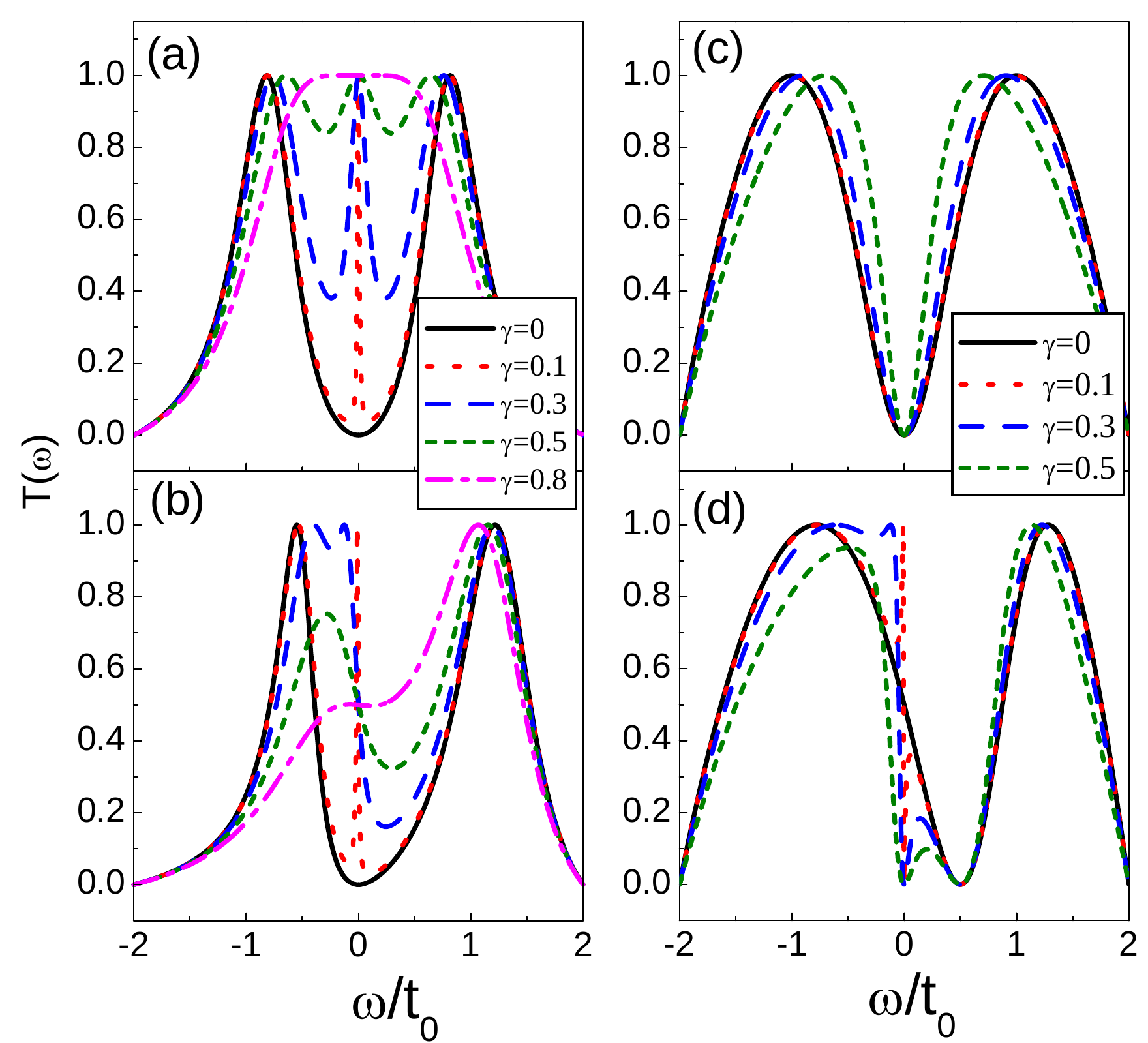}}
\caption{Spectra of the transmission function of the triple-QD chain influenced by the $\mathcal{PT}$-symmetric complex potentials, in the two cases of $v_1=0$ and $v_2=0$, respectively. (a)-(b) Transmission function spectra of $\gamma$=0, 0.1, 0.3, 0.5, 0.8, in the cases of $E_2=0$ and $E_2=0.5$ when $v_1=0$. (c)-(d) Results of $\gamma$=0, 0.1, 0.3, 0.5, in the cases of $E_2=0$ and $E_2=0.5$ when $v_2=0$.  \label{case11}}
\end{figure}

\section{Numerical results and discussions \label{result2}}
Following the theory in the above section, we proceed to investigate the transmission function spectra of our considered system to clarify the influence of the $\mathcal{PT}$-symmetric complex potentials on the quantum transport process. In order to satisfy the condition of $\cal PT$ symmetry, we choose $t_{1(2)}=t_c$, $v_{\alpha 1(3)}=v_1$, and $v_{\alpha 2}=v_2$ with $E_{1(3)}=E_0\pm i\gamma$. Besides, we take $E_0=0$ and assume $t_0$ to be the unit of energy for calculation. In the context, we would like to pay attention to two cases, i.e., triple-QD chain and triple-QD ring, to expand our numerical discussion.
\begin{figure}[htb]
\centering \scalebox{0.36}{\includegraphics{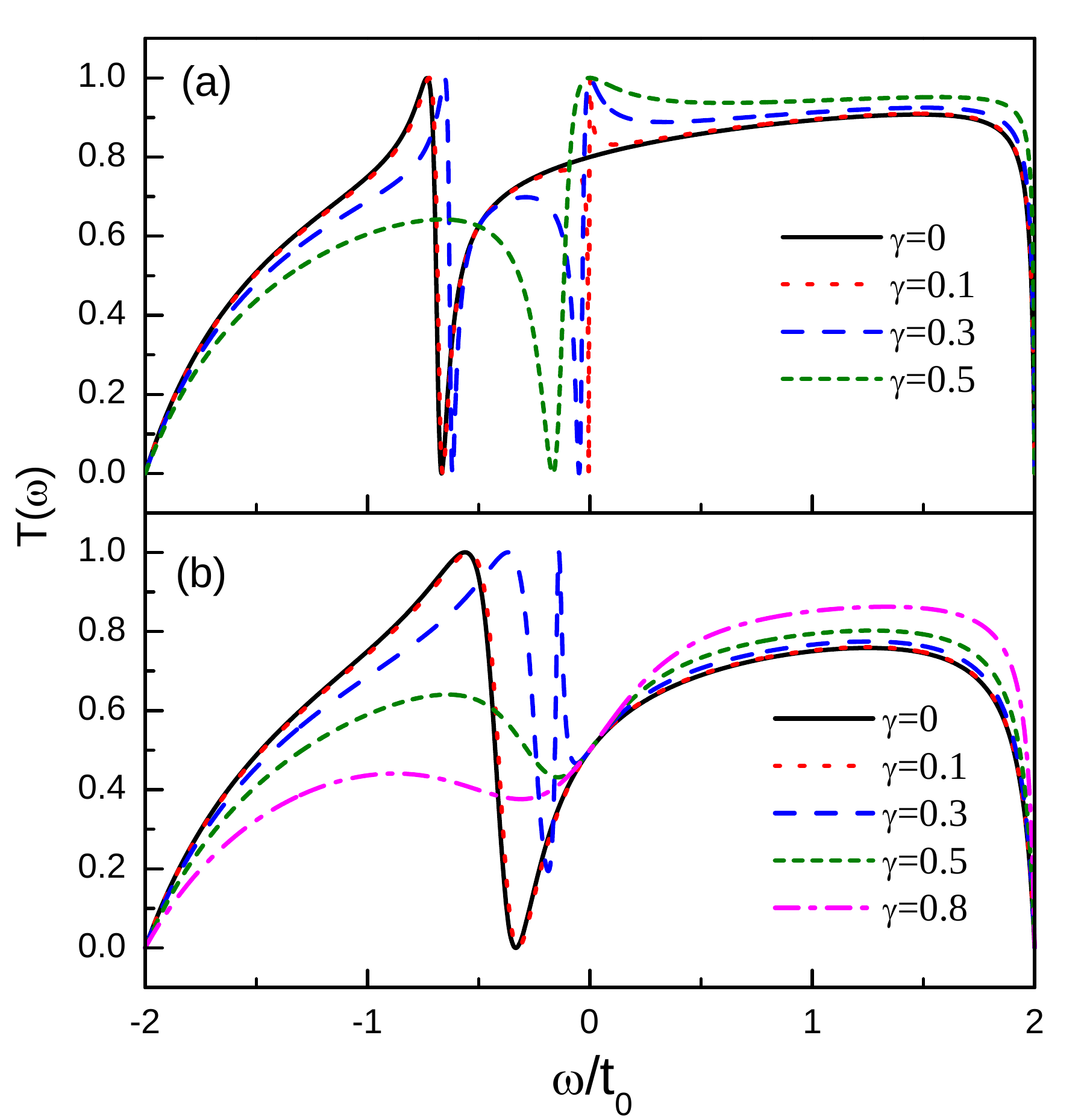}} \caption{
Spectra of $T(\omega)$ with the increase of $\gamma$. The QD-lead couplings are taken to be $v_1=v_2=0.5$. (a) Transmission function spectra of $\gamma$=0, 0.1, 0.3, 0.5, in the case of $E_2=0$. (b) Results of $\gamma$=0, 0.1, 0.3, 0.5, 0.8, when $E_2=0.5$.\label{case12}}
\end{figure}
\par
\subsection{Triple-QD chain}
In the first subsection, we would like to investigate the transmission function in the case of the triple-QD chain, by considering different QD-lead coupling manners. The numerical results are shown in Fig.\ref{case11}.
In Fig.\ref{case11}(a)-(b), we take $v_1=0$ and $v_2=0.5$ and investigate the quantum transport properties in the simplest geometry, i.e., the cross-typed triple QDs. For performing the numerical calculation, we take the interdot coupling as $t_{c}=0.5$ without loss of generality. In Fig.\ref{case11}(a), we see that for the Hermitian Hamiltonian of the triple-QD chain (i.e., $\gamma=0$), only two peaks appear in the transmission function spectrum, near the positions of $\omega\approx\pm 1.0$ respectively. Also, the two peaks are separated from each other by one antiresonance at the energy zero point. Thus, in this case, the decoupling mechanism occurs in the quantum transport process, accompanied by the occurrence of antiresonance. As the non-Hermitian Hamiltonian is taken into account by setting $E_1=E_0-i\gamma$ and $E_3=E_0+i\gamma$, one can readily find that both the decoupling and antiresonance vanishes. It shows that even if $\gamma=0.1$, the antiresonance disappears, whereas one resonant peak emerges at the energy zero point. If the value of $\gamma$ further increases, all the peaks in the transmission function spectra are widened obviously. As a result, in the case of $\gamma=0.8$, one transmission function plateau forms and $T(\omega)$ gets close to $1.0$ around the position of $\omega=0$.
Fig.\ref{case11}(b) shows one general case where the level of QD-2 is away from the energy zero point, e.g., $E_2=0.5$. It can be found that in the case of $\gamma=0$, the decoupling and antiresonance phenomena still co-exist in the quantum transport process, though the two peaks shifts to the positions of $\omega=-0.7$ and $\omega=1.2$, with their different widthes. Next, when the nonzero $\gamma$ is taken into account, both the decoupling and antiresonance disappear, similar to the case of $E_2=0$. For a small $\gamma$ (i.e., $\gamma=0.1$), one resonant peak emerges at the energy zero point. With its increase, all the peaks in the transmission function spectra are widened. At the same time, the transmission functions in different cases are equal to one another around the position of $\omega=0$, with $T(\omega)\approx 0.5$. As a result, in the negative region the transmission-function peak can be suppressed until its disappearance.
\par
Next, we focus on an alternative case where $v_1=0.5$ and $v_2=0$, with the results exhibited in Fig.\ref{case11}(c)-(d).
We find in Fig.\ref{case11}(c) that in the case of $E_l=E_0$, two peaks exist in the transmission function curve, and at the point of $\omega=E_0$ the transmission function encounters its zero. Such a result is similar to that in Fig.\ref{case11}(a), except for the difference of the peak widths. However, the influence of the $\mathcal{PT}$-symmetric complex potentials on the quantum transport shows new results. Namely, the nonzero $\gamma$ only narrows the antiresonance valley around the energy zero point but does not induce any other phenomenon. Next in the case of $\delta=0.5$, Fig.\ref{case11}(d) shows that the antiresonance point shifts to the position of $\omega=0.5$. As the nonzero $\gamma$ is taken into account, it causes a new antiresonance to appear at the energy zero point, with the antiresonance valley proportional to the value of $\gamma$. As a consequence, two antiresonance points appear in the transmission function spectrum. Up to now, we can find that the effect of the $\mathcal{PT}$-symmetric complex potentials is strongly dependent on the QD-lead coupling manner.
\par
One can be sure that the spectra properties of the transmission function can be understood by writing out its analytical expression. To do so, we rewrite $T(\omega)$ in a form as $T(\omega)=|\tau_t|^2=|\sum_{jl}\tilde{v}_{Lj}G_{jl}\tilde{v}_{jR}|^2$ with $\tilde{v}_{\alpha j}=v_{\alpha j}\sqrt{\rho_0}$. For the general case, we can obtain the result that
\begin{widetext}
\begin{eqnarray}
\tau_t={2\Gamma_1(\omega-E_2)(\omega-E_0)+\Gamma_2[(\omega-E_0)^2+\gamma^2]+4\sqrt{\Gamma_1\Gamma_2}(\omega-E_0)t_c\over \prod_j(\omega-E_j-\Sigma_{jj})-2{\textrm{Re}}[(t_c+\Sigma_{2 1})(t_c+\Sigma_{32})\Sigma_{13}]-{\cal D}}\label{transaa}
\end{eqnarray}
\end{widetext}
where ${\cal D}=|t_c+\Sigma_{21}|^2(\omega-E_3-\Sigma_{33})
+|\Sigma_{13}|^2(\omega-E_2-\Sigma_{22})
+|t_c+\Sigma_{32}|^2(\omega-E_1-\Sigma_{11})$ with $\Gamma_n=v_n^2\rho_0$ ($n=1,2$). One can find the condition of $\tau_t=0$, i.e.,
\begin{eqnarray}
\omega=E_0-{2\over 2+x}(2\sqrt{x}t_c-\delta\pm\sqrt{\Delta})
\end{eqnarray}
in which $\Delta=(\delta-2\sqrt{x}t_c)^2-x(x+2)\gamma^2$ with $x={\Gamma_2\over\Gamma_1}$. In the case of $v_1=0$, $\tau_t$ will simplify to be
$\tau_t={\Gamma_2\over \omega-E_2-\Sigma_{22}-2{\omega-E_0\over(\omega-E_0)^2+\gamma^2}t_c^2}$. This result exactly means that the presence of nonzero $\gamma$ will destroy the antiresonance at the point of $\omega=E_0$. When $\omega=E_0$, $\tau_t$ will be simplified into $\tau_t={\Gamma_2\over \omega-E_2-\Sigma_{22}}$. On the other hand, in the case of $v_2=0$, $\tau_t$ will transform into
$\tau_t={2\Gamma_1(\omega-E_2)(\omega-E_0)\over {\cal D}'(\omega-E_2)-2t_c^2(\omega-E_0)}$
with ${\cal D}'=(\omega-E_0)^2+\gamma^2-(\Sigma_{11}+\Sigma_{33})(\omega-E_0)$. Thus in such case, the antiresonance will occur at the points of $\omega=E_2$ and $\omega=E_0$, once $\gamma$ is not equal to zero. Alternatively, in the absence of the $\cal PT$-symmetric complex potentials, $\tau_t$ will be simplified to be $\tau_t={2\Gamma_1(\omega-E_2)\over (\omega-E_0-t_3-\Sigma_{11}+\Sigma_{33})(\omega-E_2)-2t_c^2}$, which suggests only one antiresonance point located at the position of $\omega=E_2$.

\par
Based on the result in Eq.(\ref{transaa}), we assume $v_1=v_2=0.5$ and investigate the effect of $\mathcal{PT}$-symmetric complex potentials on the quantum transport behaviors. The results are shown in Fig.\ref{case12}. It can be clearly found that in this case, the $\mathcal{PT}$-symmetric complex potentials take complicated effect to the quantum transport process. Firstly, in Fig.\ref{case12}(a) where $E_2=0$, we see that in the case of $\gamma=0$, an apparent Fano line shape exists in the transmission function spectrum, with the antiresonance point at $\omega\approx -0.7$. When the nonzero $\gamma$ is introduced, one new antiresonance point appears in the vicinity of the energy zero point. However, the further increase of $\gamma$ will eliminate the antiresonance near the point of $\omega\approx -0.7$, accompanied by the enhancement of the antiresonance around the energy zero point. As $\gamma$ rises to 0.5, there is only one antiresonance point at $\omega\approx -0.2$. Secondlu, in the case of $E_2=0.5$, the effect of $\mathcal{PT}$-symmetric complex potentials is only to weaken the Fano interference. As shown in Fig.\ref{case12}(b), for the small $\gamma$, i.e., $\gamma=0.1$, these potentials play trivial roles in changing the transmission function spectrum. Once $\gamma$ increases, the Fano antiresonance will be destroyed gradually. For instance, in the case of $\gamma=0.5$, the Fano lineshape in the transmission function spectrum disappears.

\begin{figure}[htb]
\centering \scalebox{0.35}{\includegraphics{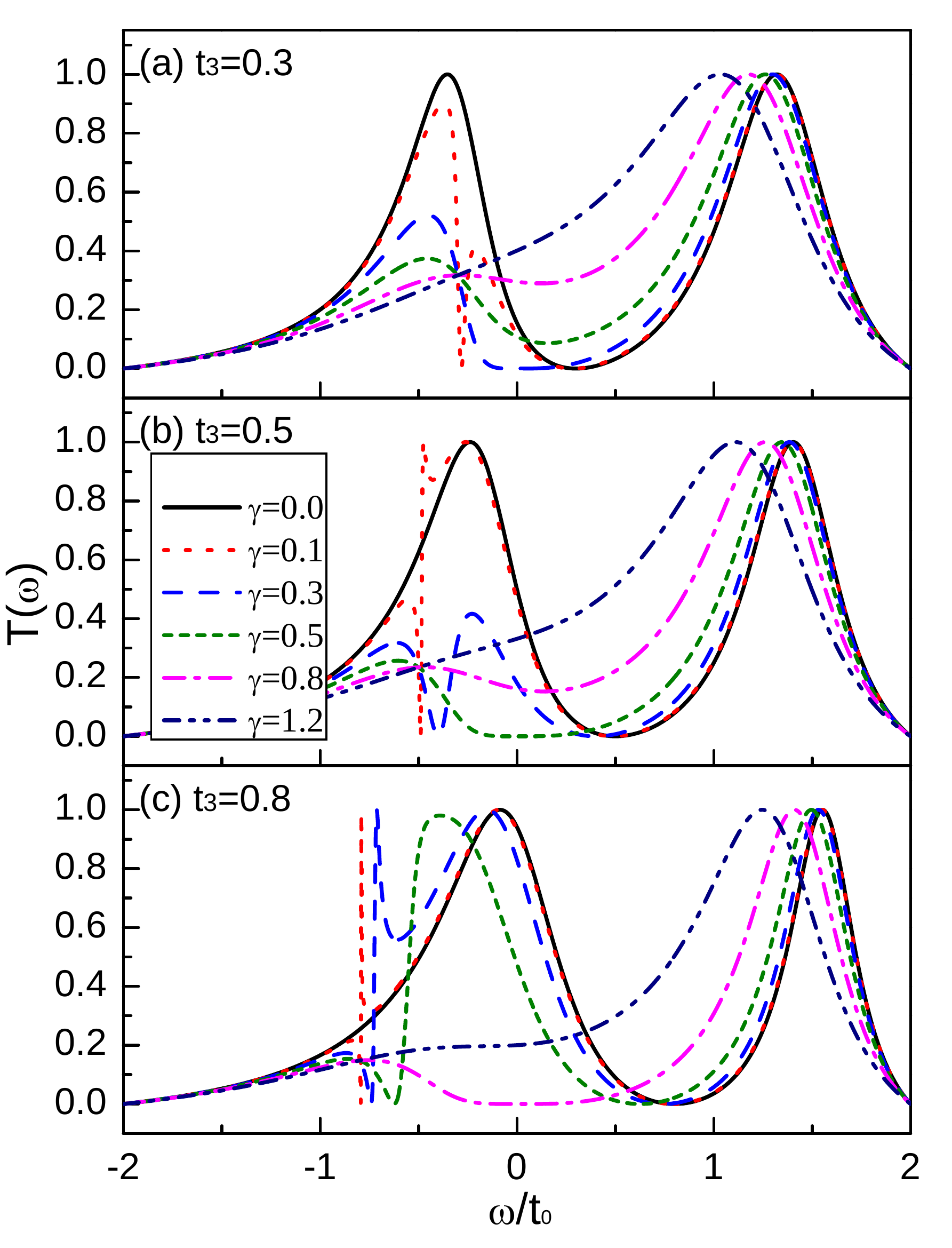}} \caption{Transmission function spectra of the triple-QD ring affected by the $\mathcal{PT}$-symmetric complex potentials, in the case of $v_1=0$ and $E_2=0.5$. In (a)-(c), $t_3$ is equal to 0.3, 0.5, and 0.8, respectively. \label{case21}}
\end{figure}
\subsection{Triple-QD ring}
\par
In what follows, we would like to introduce the coupling between QD-1 and QD-3 to investigate the transmission function properties in the triple-QD ring. Similar to the discussion in the above subsection, we will consider the QD-lead coupling manners of $v_1=0$ and $v_2=0$, respectively. In order to present a general description about the effect of the $\cal PT$-symmetric complex potentials, we take $E_2=0.5$ in this part.
\begin{figure}[htb]
\centering \scalebox{0.35}{\includegraphics{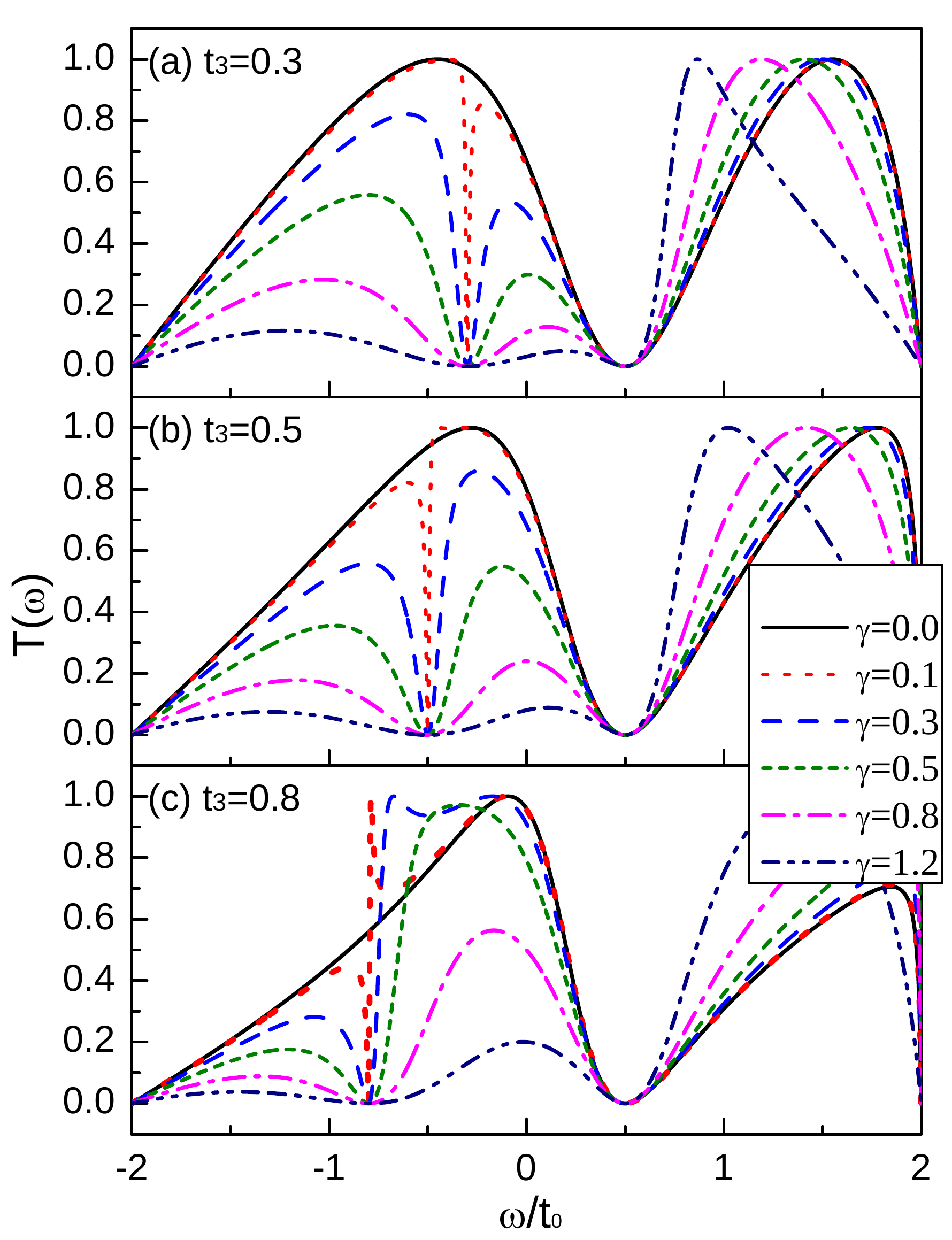}} \caption{
Transmission function spectra of the triple-QD ring in the presence of $\mathcal{PT}$-symmetric complex potentials. The QD-lead coupling is taken to $v_2=0$, and the level of QD-2 is fixed at $E_2=0.5$. In (a)-(c), $t_3$ is equal to 0.3, 0.5, and 0.8, respectively.\label{case22}}
\end{figure}
\par
The results in Fig.\ref{case21} describe the case of $v_1=0$ and $v_2=0.5$. In this figure, we see that in the case of $\gamma=0$, only two peaks appear in the transmission function spectrum, and they are separated by the antiresonance at the point of $\omega=t_3$. Next when the non-Hermitian Hamiltonian is considered with $E_1=E_0-i\gamma$ and $E_3=E_0+i\gamma$, the transmission function spectrum undergoes complicated change. It shows that the peak in the high-energy region shifts left, accompanied by its widening. For the antiresonance, one new antiresonance emerges and it is located at the position of $\omega=-t_3$ when $\gamma=0.1$. With the increase of $\gamma$, this antiresonance will be enhanced, leading to the disappearance of the transmission function peak beside this antiresonance. Meanwhile, such a new antiresonance point shifts right, until the consistence of the two antiresonances in the case of $\gamma=t_3$. Next, the further increase of $\gamma$ will destroy the antiresonance phenomenon, and then only one peak survives in the transmission function spectrum.
For the QD-lead coupling manner of $v_1=0.5$ and $v_2=0$, we see in Fig.\ref{case22} that in such a case, the presence of $\cal PT$-symmetric complex potentials leads to alternative results. The most typical result is that it introduces one antiresonance at the point of $\omega=-t_3$, whereas the original antiresonance is still located at the point of $\omega=\delta$  the transmission function spectrum. In addition, with the increase of $\gamma$, the antiresonance valley around that point of $\omega=\delta$ is narrowed, but the other one is widened. As a consequence, the transmission function is suppressed in this process. For instance, in the case of $\gamma=1.2$, some of the transmission function peaks disappear.
In view of the above two cases, one can find that the effect of the complex potentials is more complicated in the first case.
\par
\begin{figure}[htb]
\centering \scalebox{0.35}{\includegraphics{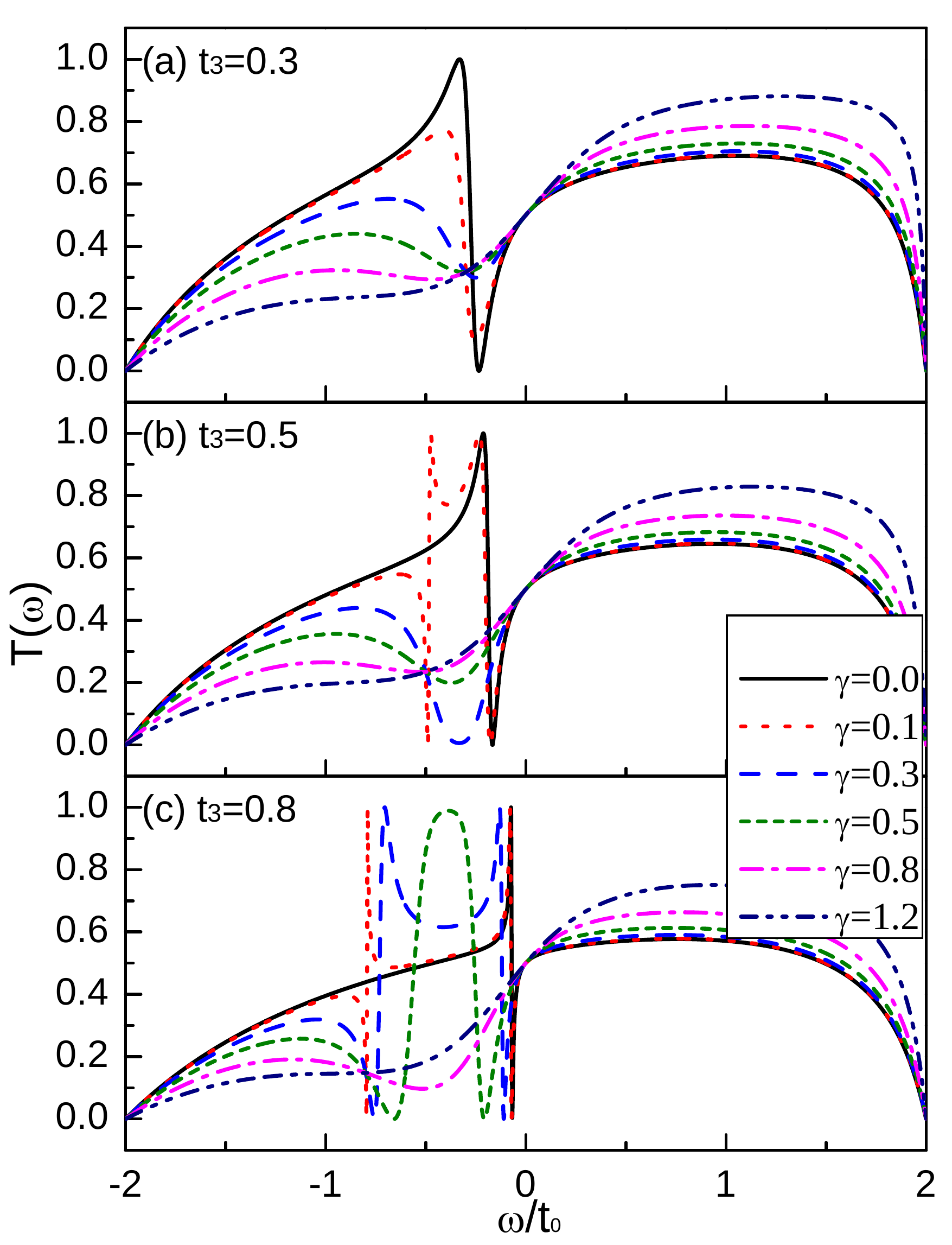}} \caption{
Spectra of $T(\omega)$ in the triple-QD ring with the increase of $\mathcal{PT}$-symmetric complex potentials. The structural parameters are $v_{1(2)}=0.5$ and $E_2=0.5$. In (a) $t_3=0.3$, $t_3=0.5$ in (b), and $t_3=0.8$ in (c). \label{case23}}
\end{figure}

In Fig.\ref{case23}, we suppose $v_1=v_2=0.5$ to analyze the influence of ${\cal PT}$-symmetric complex potentials. In this figure, we can find that the influence of $\cal PT$-symmetric complex potentials is dependent on the value of $t_3$. To be specific, in the case of $t_3=0.3$, the nonzero $\gamma$ can efficiently destroy the Fano antiresonance in the transmission function spectrum. With the increase of $\gamma$, the Fano lineshape in the transmission function spectrum disappears gradually. When $t_3$ increases to 0.5, one new antiresonance point appears at the point of $\omega=-0.5$ in the case of $\gamma=0.1$. However, as $\gamma$ increases to 0.3, the two antiresonance change to one, at the point of $\omega=-0.3$. The following increase of $\gamma$ will eliminate this antiresonance and destroy the Fano lineshape in the transmission function spectra.
In the case of $t_3=0.8$, the influence of increasing $\gamma$ is similar to the result in Fig.\ref{case23}(b). The difference consists in that in the case of $\gamma\ge 0.8$, the Fano antiresonance vanishes.
\par
Let us present the analytical expression of the transmission coefficient in the case of $t_3\neq 0$. In the general case where $v_{1(2)}\neq0$, it is given by
\begin{widetext}
\begin{eqnarray}
\tau_t={2\Gamma_1(\omega-E_2)(\omega-E_0+t_3)+\Gamma_2[(\omega-E_0)^2+\gamma^2-t^2_3]+4\sqrt{\Gamma_1\Gamma_2}[(\omega-E_0)+t_3]t_c\over \prod_j(\omega-E_j-\Sigma_{jj})-2{\textrm{Re}}[(t_c+\Sigma_{2 1})(t_c+\Sigma_{32})(t_3+\Sigma_{13})]-{\cal D}'}
\end{eqnarray}
\end{widetext}
where ${\cal D}'=|t_c+\Sigma_{21}|^2(\omega-E_3-\Sigma_{33})
+|t_3+\Sigma_{13}|^2(\omega-E_2-\Sigma_{22})
+|t_c+\Sigma_{32}|^2(\omega-E_1-\Sigma_{11})$ with $\Gamma_n=v_n^2\rho_0$ ($n=1,2$). One can find the condition of $\tau_t=0$, i.e.,
\begin{eqnarray}
\omega=E_0-{2\over 2+x}(t_3-\delta+2\sqrt{x}t_c\pm\sqrt{\Delta})
\end{eqnarray}
in which $\Delta=(t_3-\delta+2\sqrt{x}t_c)^2-(x+2)[x(\gamma^2-t_3^2)-2t_3(\delta-2\sqrt{x}t_c)]$ with $x={\Gamma_2\over\Gamma_1}$. Here, the role of $t_3$ can be clearly observed. It also shows that with the increase of $\gamma$, $\Delta$ has an opportunity to be equal to or less than zero, consequently, the antiresonance points will decrease and then disappear. Take the case of $t_3=t_c$ and $x=1$ as an example, antiresonance will disappear when $\gamma>{\delta\over\sqrt{3}}$. Next, when our considered structure is simplified, the antiresonance properties will become clearer accordingly.
In the case of $v_1=0$, $\tau_t$ will simplify to be
\begin{eqnarray}
&&\tau_t={\Gamma_2\over \omega-E_2-\Sigma_{22}-2{\omega-E_0+t_3\over(\omega-E_0)^2+\gamma^2-t^2_3}t_c^2}.
\end{eqnarray}
From this result, we can find that the antiresonance occur at the point of $\omega=E_0\pm\sqrt{t_3^2-\gamma^2}$. However, in the case of $\gamma=0$,
$\tau_t={\Gamma_2\over \omega-E_2-\Sigma_{22}-2{t_c^2 \over\omega-E_0-t_3}}$.
Antiresonance only occurs at the point of $\omega=E_0+t_3$, whereas the antiresonance point $\omega=E_0-t_3$ disappears due to the molecular-level decoupling. On the other hand, in the case of $v_2=0$, $\tau_t$ will transform into
$\tau_t={2\Gamma_1(\omega-E_2)(\omega-E_0+t_3)\over {\cal D}'(\omega-E_2)-2t_c^2(\omega-E_0+t_3)}$
with ${\cal D}'=(\omega-E_0)^2+\gamma^2-t^2_3-(\Sigma_{11}+\Sigma_{33})(\omega-E_0+t_3)$.
This result means that the antiresonance occurs at the point of $\omega=E_2$ and $\omega=E_0-t_3$, independent of the change of nonzero $\gamma$. Next in the absence of the $\cal PT$-symmetric complex potentials, $\tau_t$ will be simplified to be  $\tau_t={2\Gamma_1(\omega-E_2)\over (\omega-E_0-t_3-\Sigma_{11}+\Sigma_{33})(\omega-E_2)-2t_c^2}$. And then, only one antiresonance point can be observed, which is located at the position of $\omega=E_2$.
\par
\begin{figure}[htb]
\centering \scalebox{0.35}{\includegraphics{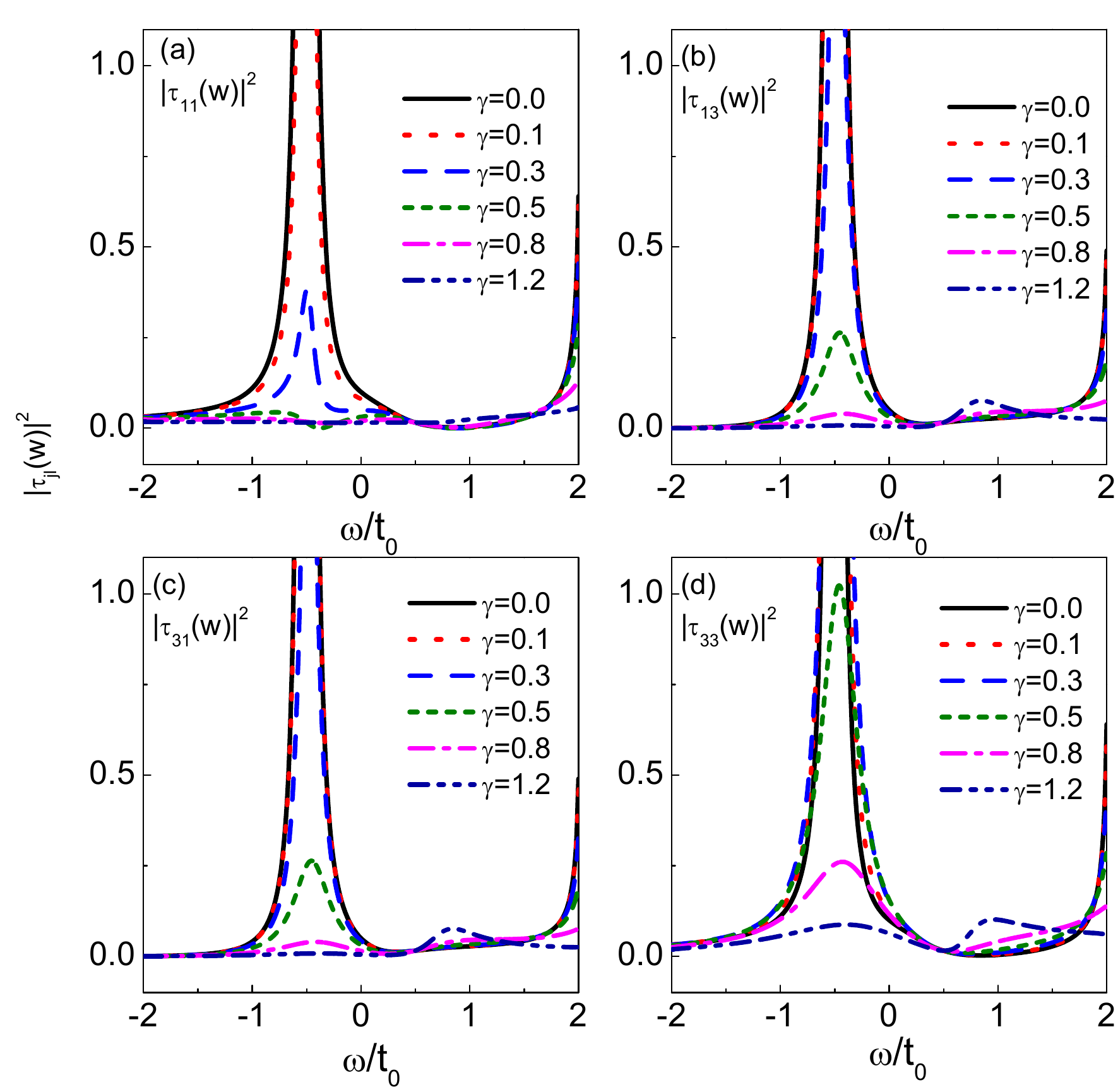}} \caption{Magnitude of $\tau_{jl}(\omega)$ in the triple-QD ring with the increase of $\mathcal{PT}$-symmetric complex potentials. Structural parameters are $t_{3}=E_2=0.5$ and $v_1=0.5$. \label{Mag}}
\end{figure}

\begin{figure}[htb]
\centering \scalebox{0.55}{\includegraphics{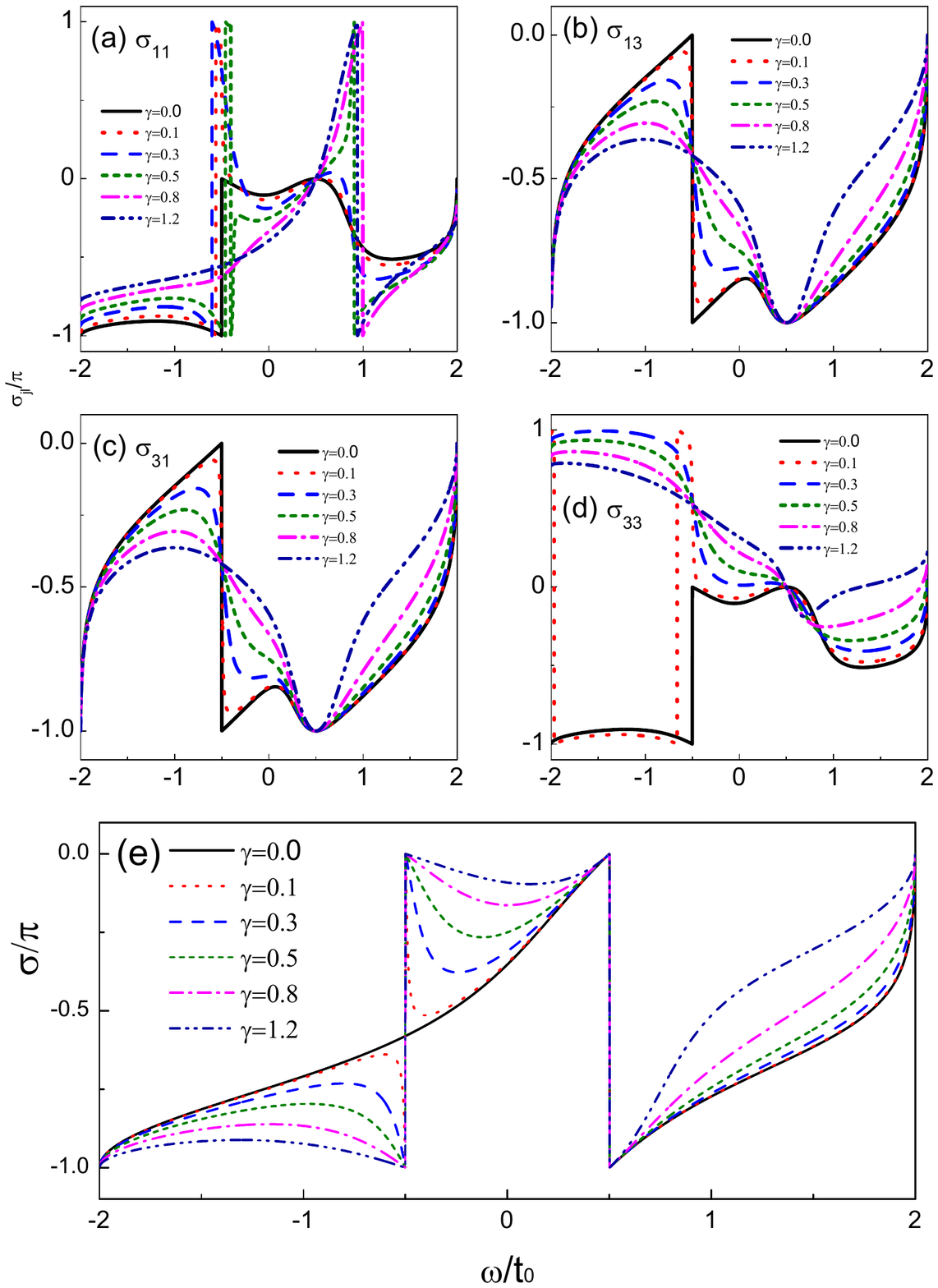}} \caption{(a)-(d) Phase of $\tau_{jl}(\omega)$ in the triple-QD ring influenced by the increase of $\gamma$. (e) Phase change of $\tau(\omega)$ in the triple-QD ring. The relevant parameters are identical with those in Fig.\ref{Mag}. \label{Phase1}}
\end{figure}

\subsection{Quantum interference analysis}
It is well known that the property of the transmission function is determined by the quantum interference among the transmission paths in the low-dimensional system. Thus, the influence of the $\cal PT$-symmetric complex potentials on the transmission function originates from its nontrivial contribution to the quantum interference. Following this idea, we here would like to discuss the quantum interference change in the presence of the $\cal PT$-symmetric complex potentials.
\par
In this subsection, we take the case of triple-QD ring with $v_1=0.5$ and $v_2=0$ as an example to expand discussion. As demonstrated in the above subsection, the transmission coefficient $\tau_t$ obeys the following relationship, i.e., $\tau_t=\tau_{11}+\tau_{13}+\tau_{31}+\tau_{33}$. Hence, the interference among these four transmission paths governs the leading property of the transmission function. In order to clarify the interference property, in Figs.\ref{Mag}-\ref{Phase1} we plot the magnitudes and phases of these four paths affected by the $\cal PT$-symmetric complex potentials. It can be found in Fig.\ref{Mag} that such complex potentials indeed change the magnitudes of the respective transmission paths. In the vicinity of $\omega=-0.5$, the magnitudes of these transmission paths are apparently suppressed by the increase of $\gamma$, especially for $|\tau_{11}|^2$. Next, in Fig.\ref{Phase1}(a)-(d) it shows that the phases of these four paths are also varied by the complex potentials. The nontrivial change is manifested as the smoothness of the phase transition near the point of $\omega=-0.5$.
All these results inevitably modify the quantum interference in this structure. And then, we investigate the phase of the transmission coefficient, with the results shown in Fig.\ref{Phase1}(e). It can be observed that in the presence of complex potentials, the phase of $\tau_t$ experiences new $\pi$-phase jump when the incident-particle energy is tuned to $\omega=-0.5$. This should be attributed to the enhancement of the destructive quantum interference among the four transmission paths when the complex potentials are taken into account. Surely, such a jump is exactly consistent with the antiresonance point in transmission-function spectrum. Therefore, the complex potentials re-regulate the transmission function by changing the quantum interference mechanism in this system.

\begin{figure}[htb]
\centering \scalebox{0.37}{\includegraphics{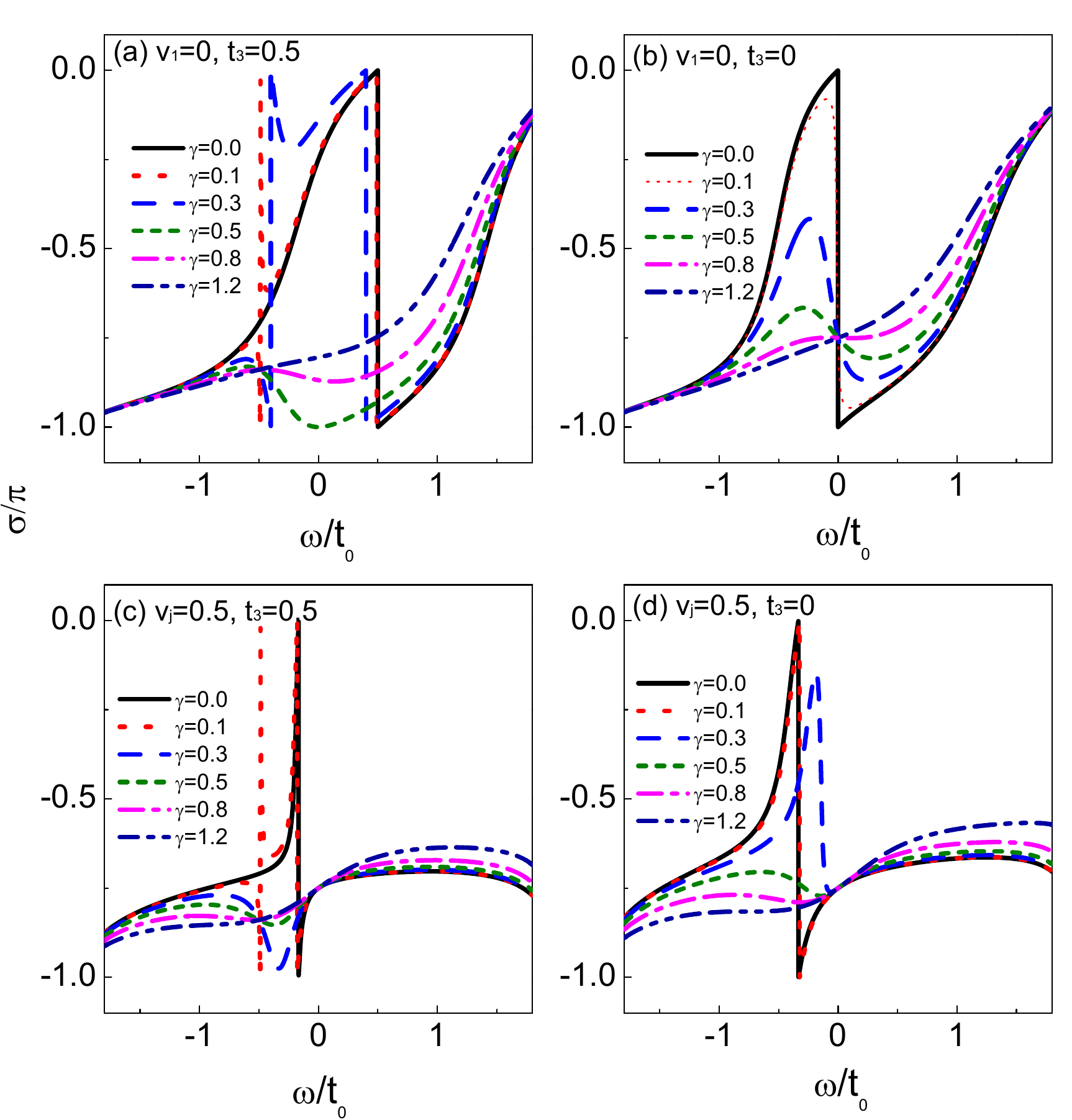}} \caption{Phases of $\tau_t(\omega)$ in the triple-QD ring and chain with the increase of $\mathcal{PT}$-symmetric complex potentials. (a)-(b) Results of $v_1=0$ and $v_2=0.5$ in the cases of $t_3=0.5$ and $t_3=0$, respectively. (c)-(d) Results of $v_j=0.5$ when $t_3=0.5$ and $t_3=0$, respectively. \label{PhaseL}}
\end{figure}
\par
In order to completely describe the relation between the antiresonance and the phase jump of $\tau_t$, we investigate the phases of the transmission coefficients in the other cases, with the numerical results shown in Fig.\ref{PhaseL}. Fig.\ref{PhaseL}(a)-(b) exhibits the phases of $\tau_t$ in the structures of $t_3=0.5$ and $t_3=0$, respectively, under the situation of $v_1=0$ and $v_2=0.5$. It can be found that increasing $\gamma$ first induces a new phase jump and then removes all the phase jump in the case of $t_3=0.5$. Instead, in the case of $t_3=0$, $\gamma$ is able to remove the phase jump directly. Similar results can be observed in the case of $v_j=0.5$, as shown in Fig.\ref{PhaseL}(c)-(d). When comparing the magnitudes and phases of the transmission coefficients, one can readily find that regardless of the geometry change of the triple-QD structure, the jump of the transmission-coefficient phase is exactly consistent with the antiresonance point in the transmission-function spectrum. Up to now, the quantum interference that dominates the transport process can be understood.

\section{summary\label{summary}}
To sum up, we have presented an analysis about the effect of $\mathcal{PT}$-symmetric complex potentials on the transport properties of non-Hermitian systems, which is formed by the coupling between a triple-QD molecule and two semi-infinite leads. By analytically solving the scattering process, we have found that the $\mathcal{PT}$-symmetric imaginary potentials can induce pronounced effects on transport properties of our systems, including changes from antiresonance to resonance, shift of antiresonance, and occurrence of new antiresonance, which are related to the QD-lead coupling manner. All these results have been discussed by analyzing the quantum interference properties in the presence of the complex potentials. Our study provides an additional way to understand the physical interplay between the quantum transport and $\mathcal{PT}$ symmetry in non-Hermitian discrete systems.
%\section*{Acknowledgments}

\clearpage
%\section{\protect\bigskip\ {\protect\large FIGURES}}

\bigskip

\end{document}